\documentclass[final,5p,times,floatfix]{elsarticle}

\usepackage{graphicx,hyperref}
\newcommand{\mr}[1]{{\mathrm{#1}}}

\usepackage{amssymb,color,amsmath,bm,mathrsfs}

\usepackage[T1]{fontenc}
\usepackage{listings}
\def\code{\@ifnextchar[{\@with}{\@without}}%
\def\@with[#1]#2{%
}
\def\@without#1{%
  \section{\protect\detokenize{#1}}%
  \lstinputlisting[]{#1}%
}

\definecolor{greencode}{RGB}{0,128,0}
\definecolor{comment}{RGB}{128,128,128}
\definecolor{orange}{RGB}{255,127,0}
\newcounter{bla}

\usepackage{etoolbox,calc}

\makeatletter
\def\appendixname{Appendix}
\appto\appendix{%
  \addtocontents{toc}{\patch@l@section}
  \appto\appendixname{ }
}
\protected\def\patch@l@section{%
  \patchcmd{\l@section}{1.5em}{\widthof{\appendixname\space}+2.5em}{}{}%
}
\makeatother

\journal{Computer Physics Communications}

\begin{document}

\begin{frontmatter}



\title{ARC: An open-source library for calculating properties of alkali Rydberg atoms}


\author[a]{N. \v{S}ibali\'{c}\corref{author}}
\author[b]{J. D. Pritchard}
\author[a]{C. S. Adams}
\author[a]{K. J. Weatherill}

\cortext[author] {Corresponding author.\\\textit{E-mail address:} nikolasibalic@physics.org}
\address[a]{Joint Quantum Center (JQC) Durham-Newcastle, Department of Physics, Durham University, South Road, Durham, DH1 3LE, United Kingdom}
\address[b]{Department of Physics, SUPA, University of Strathclyde, 107 Rottenrow East, Glasgow, G4 0NG, United Kingdom}

\begin{abstract}

We present an object-oriented Python library for computation of properties of highly-excited Rydberg states of alkali atoms. These include single-body effects such as dipole matrix elements, excited-state lifetimes (radiative and black-body limited) and Stark maps of atoms in external electric fields, as well as two-atom interaction potentials accounting for dipole and quadrupole coupling effects valid at both long and short range for arbitrary placement of the atomic dipoles. The package is cross-referenced to precise measurements of atomic energy levels and features extensive documentation to facilitate rapid upgrade or expansion by users.  This library has direct application in the field of quantum information and quantum optics which exploit the strong Rydberg dipolar interactions for two-qubit gates, robust atom-light interfaces and simulating quantum many-body physics, as well as the field of metrology using Rydberg atoms as precise microwave electrometers.

\end{abstract}


\end{frontmatter}


{\bf PROGRAM SUMMARY}

\begin{small}
\noindent
{\em Program Title:} ARC: Alkali Rydberg Calculator \\
{\em Licensing provisions:} BSD-3-Clause\\
{\em Programming language:} Python 2.7 or 3.5, with C extension\\
{\em Computer:} i386, x86-64 \\
{\em Operating System:} Linux, Mac OSX, Windows\\
{\em RAM:} of the order of several 100MB for calculations involving several 1000's of basis states \\
{\em External Routines:} NumPy \cite{1}, SciPy \cite{1}, Matplotlib  \cite{2} \\
{\em Nature of problem:} \\ 
Calculating atomic properties of alkali atoms including lifetimes, energies, Stark shifts and dipole-dipole interaction strengths using matrix elements evaluated from radial wavefunctions.\\
{\em Solution method:}\\ 
Numerical integration of radial Schr\"odinger equation to obtain atomic wavefunctions, which are then used to evaluate dipole matrix elements. Properties are calculated using second order perturbation theory or exact diagonalisation of the interaction Hamiltonian, yielding results valid even at large external fields or small interatomic separation.\\
{\em Restrictions:}\\
External electric field fixed to be parallel to quantisation axis.\\
{\em Supplementary material:} Detailed documentation (.html), and Jupyter notebook with examples and benchmarking runs (.html and .ipynb).

\end{small}


\lstset{
  basicstyle=\ttfamily,
  columns=fullflexible,
  keepspaces=true,
}

\section{Introduction}

Highly-excited atoms, in so called Rydberg states, provide strong atom-atom interactions, and large optical nonlinearities. They are a flourishing field for quantum information processing~\cite{Paredes-Barato2014,Saffman2016a} and quantum optics~\cite{Pritchard2013,Firstenberg2016,Murray2016a} in the few to single excitation regime, as well as many-body physics \cite{Ates2007,Hoening2014a,Lesanovsky2014a,Urvoy2015,Schaus2015,Sibalic2016}, in the many-excitations limit. Their exploration requires detailed knowledge of both the single-atom properties, such as lifetimes, energies and transition dipoles elements, as well as atom pair properties, such as their interactions and strongly perturbed energy levels for atoms at distances in the micrometer range. Rydberg states are also highly sensitive to external DC and AC fields, making them excellent candidates for precision electrometry and imaging in the microwave~\cite{Sedlacek2012,Simons2016} and terahertz~\cite{Wade2016a} range, as well as performing state engineering to tune pair-state interaction potentials \cite{Ryabtsev2010,Tanasittikosol2011a,Sevincli2014}.

Although many of the relevant calculations share the same primitives, such as numeric integration of atomic wavefunctions based on measured energy levels and model core potentials, these basic efforts have been repeated by many groups independently so far. To date no single common resource has emerged for building upon more complex calculations, or for performing quick numerical estimates. To this end, we have developed Alkali Rydberg Calculator (ARC)~\cite{arcGitHub},  a library of routines written in Python, using object-oriented programming to make a modular, reusable and extendable collection of routines and data for performing useful calculations of single-atom and two-atom properties, like level diagrams, interactions and transition strengths for alkali-metal atoms.

The hierarchical nature of the package helps organise possible calculations into abstraction levels, allowing one to pick the information at the relevant level for the calculation at hand. The data for individual atomic species is provided as classes that inherit calculation methods defined as abstract classes, allowing one easily to check and update relevant data, should future measurements improve some of the experimentally estimated values. Detailed documentation is provided for all the ARC's modules~\cite{arcDocumentation2016}. In addition, the code is commented, cross-referenced in-line and uses self-descriptive names. Whenever possible, the class and function names reflect the hierarchical structure of atomic physics knowledge and the natural decompositions of the calculations, not the low-level implementation details. In addition to the documentation, ARC has example snippets provided as an IPython notebook~\cite{Sibalic2016c}, giving an overview of Rydberg physics and how to perform calculations using the package library. This is a good starting point for new users.

To facilitate the initial adoption of the package and to allow quick calculations useful in the research planning stages, we are also providing a web-interface to basic functions of the package~\cite{atomcalcWebPage}. This allows any device with a web browser to access the web-server, that will use the ARC package to perform the calculations and obtain results that are transferred back to the users' web browser. In the process, the web service self-generates and outputs the code, so it can be used as an example-on-demand service, providing a starting point for more complex calculations.

This paper is organised as follows. An overview of the ARC architecture is presented in Sec.~\ref{sec:arch}, where we introduce the theoretical framework for performing Rydberg state calculations, e.g. calculating wavefunctions and diagonalising interaction Hamiltonians. Here we also provide illustrative examples for building up calculations and visualising results with the provided tools. Initial setup for the ARC library is presented in Sec.~\ref{sec:getting_started}, and specific details relating to the implementation are discussed in Sec.~\ref{sec:implementation}. Finally, Sec.~\ref{sec:outlook} briefly summarises the package and outlines future possible expansions for the library, with a complete list of ARC classes, methods and functions detailed in Appendix A. Detailed documentation of the module is provided in .html format in the Supplemental Material \cite{arcDocumentation2016} or available from the ARC website~\cite{atomcalcWebPage}, along with an IPython notebook~\cite{Sibalic2016c} that contains numerous additional examples and code snippets that reproduce many of the results from the literature.

\section{ARC architecture and modules}\label{sec:arch}

\begin{figure*}[t!]
\begin{center}
\includegraphics[width=0.8\textwidth]{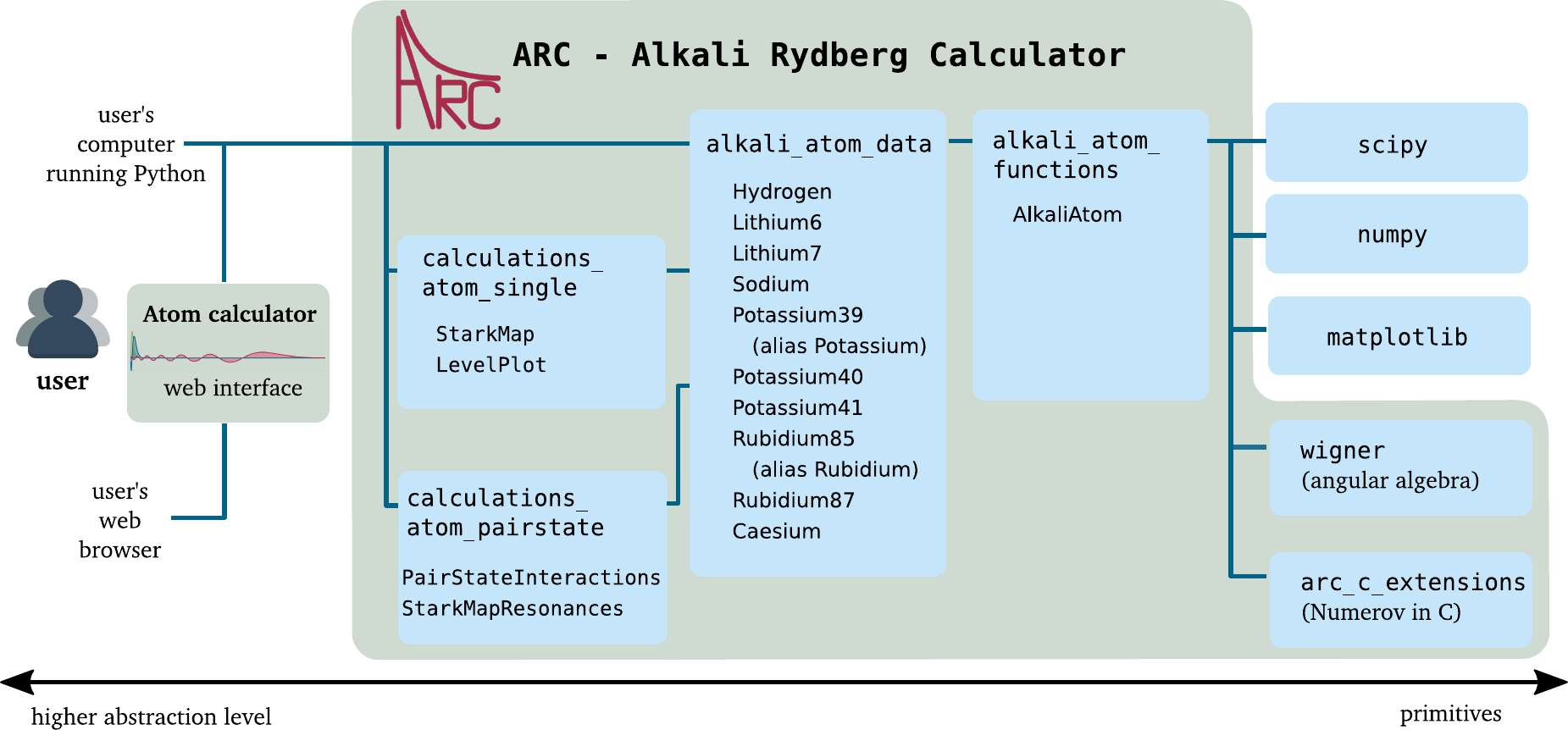}
\end{center}
\caption{\label{fig:module_overview}Overview of the Alkali Rydberg Calculator (ARC) module for Python. Object-oriented structure with hierarchy reflecting the structure of atomic physics calculations is used. This allows user to choose abstraction level at which one wants to work, from low-level implementations of basic functions of \texttt{alkali\_atom\_functions.AlkaliAtom} for finding dipole matrix elements and lifetimes, to high-level functions that allow automatic F\"orster resonance finding and exploration of complex energy-level diagrams of atomic pair-states at small inter-atomic separations in \texttt{calculations\_atom\_pairstate}.}
\end{figure*}

\subsection{Overview}

The structure of ARC library is shown in Fig.~\ref{fig:module_overview}. At the lowest level, the \verb|wigner| module implements angular momentum algebra (Wigner 3-j and 6-j coefficients and the WignerD rotation matrix), and \verb|arc_c_extensions| is a Python extension coded in C  to provide fast calculation of the radial part of the atomic wavefunctions. On a higher level, \linebreak \verb|alkali_atom_functions| uses these low-level functions to \linebreak build general methods for calculating single-atom properties, which are contained within the abstract class \linebreak \verb|AlkaliAtom| that implements calculation of dipole matrix elements, transition rates, energy levels etc. The module \linebreak \verb|alkali_atom_data| defines an explicit class for each alkali element (e.g. \verb|Rubidium87(),Ceasium()|) that encodes all relevant physical parameters and inherits the calculation methods from the parent \verb|AlkaliAtom| class. These atom classes can be passed as arguments to either of the core calculation modules, \linebreak \verb|calculations_atom_single| that implements interactive energy level diagrams (\verb|LevelPlot|) and calculates Stark maps for atoms in external fields (\verb|StarkMap|), or \verb|calculations_atom_pairstate| for dealing with two-atom effects such as long-range dipole-dipole interactions. This pair-state module provides a sophisticated interface to automatically identify F\"orster resonances for atoms in weak electric fields (\verb|StarkMapResonances|) and calculate atomic interaction potentials at both long and short range including up to quadrupole-quadrupole couplings \linebreak(\verb|PairStateInteractions|). In the following we will outline the basic functionality of the ARC module, provide the theoretical framework for the various modules and give examples of relevant calculations implemented in the library.

\subsection{The \texttt{AlkaliAtom} class}

Almost all calculated quantities in the ARC library can be derived from knowledge of the Rydberg state energy levels and matrix elements. The functions to calculate these values along with other primitive single-atom properties are encapsulated within the methods of the abstract class \verb|AlkaliAtom| defined by the module \verb|alkali_atom_functions|. This class is used in \linebreak\verb|alkali_atom_data| where each alkali-metal atom \linebreak(\verb|Lithium6|, \verb|Lithium7|, \verb|Sodium|, \verb|Potassium39|, \verb|Potassium40|, \verb|Potassium41|, \verb|Rubidium85|, \verb|Rubidium87| and \verb|Caesium|, as well as \verb|Hydrogen|) inherits calculation methods from this abstract class, and specifies all the necessary numerical values for a given atom. Calculations with the ARC library begin by initiating an atom class associated with the relevant atom, as shown in the example below for caesium.
\begin{lstlisting}
from arc import *       # initialise ARC library
atom = Caesium()      # create atom Object
\end{lstlisting}
\noindent Physical properties can then be determined using the methods in the form \verb|atom.functionName(parameters)|, where a complete list of available methods is listed in Table~\ref{tab:alkaliatom} and documentation~\cite{arcDocumentation2016}. The sections below outline the key properties.

\subsubsection{Rydberg atom energy levels}
Energy of the Rydberg state with principal quantum number $n$ and orbital and total angular momentum $\ell$ and $j$ respectively, are calculated from the Rydberg formula
\begin{equation}
E_{n,\ell,j} = E_\mr{IP} - \frac{\mr{Ry}}{(n-\delta_{n,\ell,j})^2},
\end{equation}
where $E_\mr{IP}$ is the ionisation energy threshold (\verb|ionisationEnergy|), $\mr{Ry}$ is the Rydberg constant corrected for the reduced mass \linebreak(\verb|scaledRydbergConstant|), $\mr{Ry}=m_e/(m+m_e)\mr{Ry}_\infty$ and $\delta_{n,\ell,j}$ is the quantum defect (\verb|getQuantumDefect|) given by
\begin{equation}
\delta_{n,\ell,j} = \delta_0 + \frac{\delta_2}{(n-\delta_0)^{2}}+ \frac{\delta_4}{(n-\delta_0)^{4}}+\ldots,
\end{equation}

\noindent where $\delta_0, \delta_2, \ldots$ are modified Rydberg-Ritz coefficients obtained by fitting precise measurements of the atomic energy levels~\cite{Lorenzen1983}.

Energies $E$ for a given Rydberg state relative to the ionisation limit can be obtained using \verb|getEnergy| (in eV), as well as methods \verb|getTransitionWavelength| and \linebreak\verb|getTransitionFrequency| to return the wavelength (m) or frequency (Hz) of a transition between two Rydberg states. Energies and transitions are given relative to the centres of gravity of the hyperfine split ground and excited states.

\subsubsection{Rydberg atom wavefunctions}
In order to calculate matrix elements for electric dipole and quadrupole couplings of the Rydberg states, it is necessary to calculate the radial wavefunctions $R(r)$ by numerically integrating the Radial Schr\"{o}dinger equation for valence electron. Using the substitution $\rho(r)=rR(r)$, we can remove the first order differential to obtain an equation in the form:
\begin{equation}
-\frac{1}{2\mu}\frac{\mr{d}^2 \rho}{\mr{d} r^2}+\left[\frac{\ell(\ell+1)}{2~\mu~ r^2} + V(r)\right] \rho(r)= E~\rho(r),
\label{eq:radial}
\end{equation}
where $V(r)$ is the spherically symmetric central potential \linebreak(\verb|potential|). For hydrogen, and high orbital angular momentum states $\ell>3$, this is simply the Coulomb potential of $V(r)=-1/r+V_{\rm so}(r)$, with added (relativistic) spin-orbit interaction $V_{\rm so}(r) = \alpha ~\mathbf{L}\cdot\mathbf{S}/(2r^3)$, where $\alpha$ is the fine structure constant. However for alkali atoms with closed shells it is necessary to introduce a model potential that gives a Coulomb potential at long-range and at short range accounts for effects of the core penetration of the valence electron. We adopt the core potential (\verb|corePotential|) of Marinescu {\emph{et al.}} \cite{Marinescu1994} given by the form
\begin{equation}
V(r)=-\frac{Z_\ell(r)}{r}-\frac{\alpha_\mr{c}}{2r^4}(1-e^{-(r/r_\mr{c})^6})+V_{\rm so}(r),
\end{equation}
where $\alpha_\mr{c}$ is the core polarizability, $r_\mr{c}$ is a cutoff radius introduced to avoid divergence of the polarization potential at short range and the radial charge (\verb|effectiveCharge|) is given by
\begin{equation}
Z_\ell(r)=1+(z-1)~e^{-a_1r}-r~(a_3+a_4r)~e^{-a_2r},
\end{equation}
with $\ell$-dependent parameters $a_{1..6}$ taken from Table 1 of ref. \cite{Marinescu1994} which were obtained by fitting the model potential to measured energy levels for each element.

\begin{figure}
\includegraphics[width=\columnwidth]{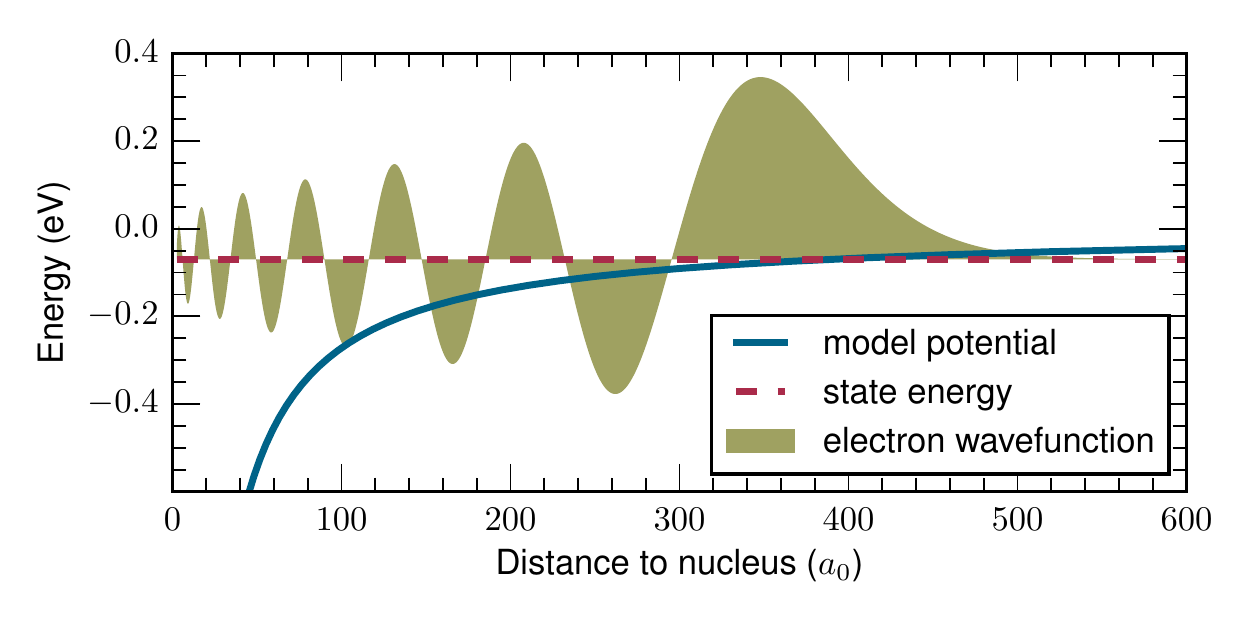}
\caption{\label{fig:wavefunction}Electron wavefunctions are calculated by Numerov integration, here exemplary shown for caesium $18~S_{1/2}m_j=1/2$ state. Solving radial part of the Schr\"odinger equation in model potential from \cite{Marinescu1994}, using state energy from quantum defects or NIST ASD database~\cite{NISTasd_ver5_4}. This calculations are starting step in calculating all atom-light couplings (dipolar and higher order).}
\end{figure}

Using this model potential and the known Rydberg energy from above, the radial wavefunctions can be calculated by numerically integrating Eq.~(\ref{eq:radial}), as shown on Fig.~\ref{fig:wavefunction}. This is achieved by performing a transformation to integrate the function $X(r)=R(r)r^{3/4}$ in terms of the scaled co-ordinate $x=\sqrt{r}$ \cite{bhatti81} that gives an approximately constant number of points across each period of oscillation in the wavefunction. The result is a second-order differential equation of the form
\begin{equation}
\frac{\mr{d}^2X}{\mr{d} x^2} = g(x)~X,
\end{equation}
which is efficiently solved using the Numerov method \cite{numerov24,numerov27}
\begin{equation}
\begin{split}
\left[1-T(x+h)\right]&~X(x+h)+\left[1-T(x-h)\right]~X(x-h) \\&= \left[2+10T(x)\right]~X(x)+O(h^6),
\end{split}
\end{equation}
where $h$ is the step size, $T(x)=h^2g(x)/12$ and
\begin{equation}
g(x) = 8~\mu~ x^2(V(r) -E)+\frac{\left(2\ell+\frac{1}{2}\right)\left(2\ell+\frac{3}{2}\right)}{x^2}.
\end{equation}

It is necessary to truncate the range of integration as at short range the model becomes unphysical and diverges, whilst at long-range the wavefunction decays to zero. Following ref.~\cite{zimmerman79}, the limits of integration are set to use an inner radius of $r_\mr{i}=\sqrt[3]{\alpha_c}$, and an outer radius of $r_\mr{o}=2n(n+15)$ which is much larger than the classical turning point of the wavefunction. To minimise errors introduced by the approximate model potential at short range, the integration is performed inwards, starting at $r_\mr{o}$. For high orbital momentum states ($\ell>3$) divergence can occur even before $r_{\rm i}$, which is automatically detected and integration is stopped at the closest wavefunction node before divergence occurred. The wavefunctions are then normalised, and returned by \verb|radialWavefunction|.

\subsubsection{Matrix elements}
Relevant properties for alkali atoms such as lifetimes \linebreak(Sec.~\ref{sec:lifetimes}), Stark shifts (Sec.~\ref{sec:stark}) module and atomic interaction potentials (Sec.~\ref{sec:interactions}) require evaluation of the electric dipole and electric quadrupole matrix elements from state $\vert n,\ell,m_\ell\rangle$ to $\vert n',\ell',m_\ell'\rangle$. For electric dipole transitions, the interaction is dependent upon matrix elements of the form $\mathcal{H}=-e\bm{r}\cdot\hat{\bm{\varepsilon}}$ where $\hat{\bm{\varepsilon}}$ is the electric field polarisation vector. Expanding the operator in the spherical basis and using the fact that the calculation can be separated into radial overlap, and angular overlap of the electron wavefunction and operator $\mathcal{H}$, the resulting matrix element can be evaluated in terms of angular momentum terms and radial matrix elements using the  Wigner-Ekart theorem \cite{sobelman79}
\begin{equation}
\langle n,\ell,m_\ell\vert r_q \vert n',\ell',m_\ell' \rangle = (-1)^{\ell-m_\ell}\begin{pmatrix}\ell&1&\ell'\\-m_\ell&-q&m_\ell'\end{pmatrix}\langle\ell \vert\vert r\vert\vert\ell'\rangle \label{eq:wignereckart},
\end{equation}
where $q$ represents the electric field polarisation ($q=\pm1,0$ driving $\sigma^\pm,\pi$ transitions respectively), and the braces denote a Wigner-3$j$ symbol (\verb|Wigner3j|). We use Condon-Shortley phase convention~\cite{Condon1970} for spherical harmonics. The reduced matrix element $\langle\ell \vert\vert r\vert\vert\ell'\rangle$ (\verb|getReducedMatrixElementL|) is\linebreak given by
\begin{equation}
\langle\ell\vert\vert r\vert\vert\ell'\rangle = (-1)^\ell\sqrt{(2\ell+1)(2\ell'+1)}\begin{pmatrix}\ell&1&\ell\\0&0&0\end{pmatrix}\mathcal{R}_{n\ell\rightarrow n'\ell'},
\end{equation}
where the radial matrix element (\verb|getRadialMatrixElement|) is evaluated from
\begin{equation}
\mathcal{R}_{n\ell\rightarrow n'\ell'} = \int_{r_{\rm i}}^{r_{\rm o}} R_{n,\ell}(r)~r~R_{n',\ell'}(r)~r^2~\mr{d} r,
\end{equation}
using numerical integration of the calculated wavefunctions.

Transforming these to the fine-structure basis Eq.~(\ref{eq:wignereckart})  can be expressed in terms of states $j,m_j$ (\verb|getDipoleMatrixElement|) as
\begin{equation}
\begin{split}
\langle n,\ell,j,m_j &\vert r_q \vert n',\ell',j',m_j' \rangle = (-1)^{j-m_j+\ell+s+j'+1}\sqrt{(2j+1)(2j'+1)}\\
&\times\begin{pmatrix}j&1&j'\\-m_j&-q&m_j'\end{pmatrix}\begin{Bmatrix}j&1&j'\\\ell'&s&\ell\end{Bmatrix}\langle \ell\vert\vert r\vert\vert\ell'\rangle, \label{eq:sphj}
\end{split}
\end{equation}
where curly braces denote a Wigner-6$j$ symbol (\verb|Wigner6j|).

To account for higher order multipole moments in the interaction between atom pairs (see Sec.~\ref{sec:interactions}), it is also necessary to calculate quadrupole matrix elements \linebreak(\verb|getQuadrupoleMatrixElement|) of the form
\begin{equation}
\mathcal{R}^Q_{n\ell\rightarrow n'\ell'} = \int_{r_{\rm i}}^{r_{\rm o}} R_{n,\ell}(r)~r^2~R_{n',\ell'}(r)~r^2\mr{d} r.
\end{equation}

As with the quantum defect model for the energies, these numeric approaches provide accurate estimates of the dipole and quadrupole terms for highly-excited states but have a large error for low-lying states where the electron density is weighted close to the atomic core where integration is most sensitive to the divergence of the model potential. To overcome this limitation, values for dipole matrix elements available in the literature either from direct measurement or more complex coupled-cluster calculations \cite{Safronova1999a,Safronova2004} are tabulated (accessible through function \verb|getLiteratureDME|). Before calculating a matrix element, the ARC library first checks if a literature value exists and if so utilises the tabulated value with the smallest estimated error. Otherwise a numerical integration is performed using the method outlined above. For each element, these tabulated values are contained within an easily readable \texttt{.csv} file (file name specified in \verb|literatureDMEfilename|) that lists the value along with relevant bibliographical reference information, making it easy for users to add new values at a later date.

\subsubsection{Rabi Frequency}
An important parameter in experiments with Rydberg atoms is the the Rabi frequency $\Omega=\bm{d}\cdot\bm{\mathcal{E}}/\hbar$ where $\bm{d}$ is the dipole matrix element for the transition and $\bm{\mathcal{E}}$ is the electric field of the laser driving the transition. For a transition from state $\vert n_1 \ell_1 j_1 m_{j_1}\rangle \rightarrow \vert n_2 \ell_2 j_2 m_{j_2}\rangle$ using a laser with power $P$ and 1/e$^2$ beam radius $w$, the Rabi frequency (in rad s$^{-1}$) can be obtained as
\begin{lstlisting}
atom.getRabiFrequency(n1,l1,j1,mj1,n2,l2,j2,mj2,P,w)
\end{lstlisting}

\noindent Related function \verb|atom.getRabiFrequency2| returns Rabi frequency based on driving electric field amplitude.

\subsubsection{Excited state lifetimes}\label{sec:lifetimes}
Radiative lifetimes of alkali atoms can be calculated using dipole matrix elements to determine the Einstein $A$-coefficient \cite{Theodesiou1984} for transitions (see \verb|getTransitionRate|),
\begin{equation}
A_{n\ell\rightarrow n'\ell'}=\frac{4\omega_{nn'}^3}{3c^3}\frac{\ell_\mr{max}}{2\ell+1}\mathcal{R}^2_{n\ell\rightarrow n'\ell'}, \label{eq:A}
\end{equation}

\noindent where $\omega_{nn'}$ is the frequency of transition from $|n\ell\rangle$ to $|n'\ell \rangle$. The total radiative decay $\Gamma_0$ is then obtained by summing over all dipole-coupled states $|n'\ell'\rangle$ of lower energy,
\begin{equation}
\Gamma_0 = \displaystyle\sum_{n\ell>n'\ell'}A_{n\ell\rightarrow n'\ell'}.
\end{equation}
At finite temperature of the environment, it is necessary to account for the interaction with black-body radiation (BBR) \cite{Beterov2009} causing stimulated emission and absorption which depend on the effective number of BBR photons per mode, $\bar{n}_\omega$, at temperature $T$ given by the Planck distribution
\begin{equation}
\bar{n}_\omega = \frac{1}{{\rm e}^{\hbar\omega_{nn'}/k_\mr{B}T}-1},
\end{equation}
multiplied by the $A$-coefficient resulting in a BBR transition rate
\begin{equation}
\Gamma_\mr{BBR}=\displaystyle\sum_{n'\ell'} A_{n\ell\rightarrow n'\ell'}~\bar{n}_{\omega_{nn'}},
\end{equation}
\noindent where summation $n',\ell'$ includes also states higher in energy, since BBR can drive these transitions. Finally, the effective lifetime $\tau$~(see \verb|getStateLifetime|) is calculated from $1/\tau = \Gamma_0+\Gamma_\mr{BBR}$.

\begin{figure}
\includegraphics[width=\columnwidth]{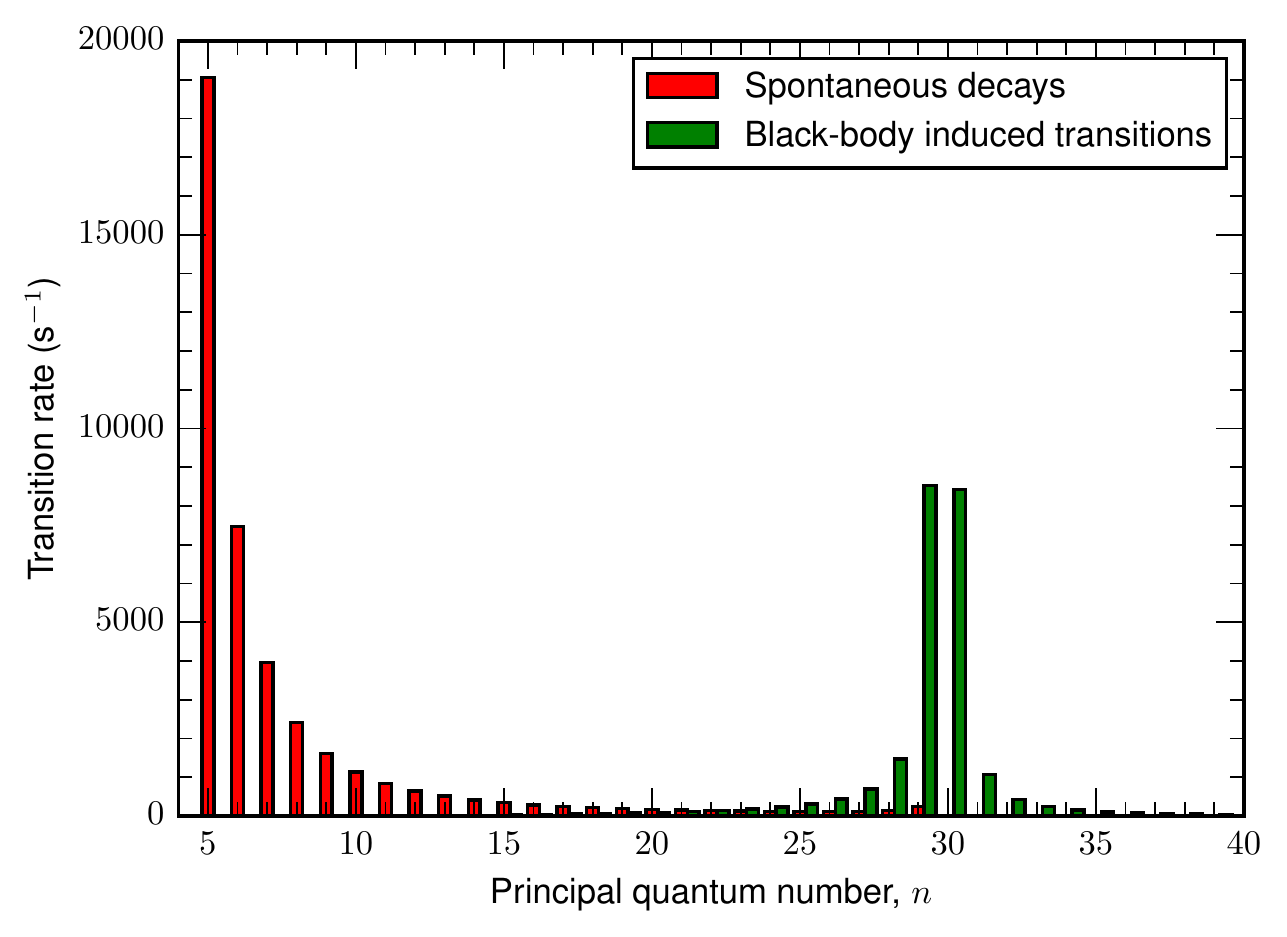}
\caption{\label{fig:bbr} Spontaneous decays and black-body induced transitions from Rb $30~S_{1/2}$ to $n~P_{1/2,3/2}$ for environment temperature of 300~K. These transitions have to be included in calculation of excited-state lifetime for Rb $30~S_{1/2}$ state.}
\end{figure}

Figure.~\ref{fig:bbr} shows the relative contribution to the excited-state lifetime of Rb $30~S_{1/2}$ due to radiative decay (dominated by low-lying states due to the $\omega^3$ scaling in Eq.~\ref{eq:A}) and black-body decay at 300~K calculated using ARC (for full code see~\ref{app:example_code}). Comparison of our calculated lifetimes~\cite{Sibalic2016c} with previous work on radiative \cite{Theodesiou1984} and BBR induced lifetimes for alkali-metals \cite{Beterov2009} yields excellent agreement.

\subsubsection{Atomic vapour properties}

For experiments using the thermal vapors of alkali atoms, a number of useful functions are provided for returning atomic vapour pressure (\verb|getPressure|) or number density \linebreak(\verb|getNumberDensity|), as well as the average interatomic distance (\verb|getAverageInteratomicSpacing|) and atomic speed \linebreak(\verb|getAverageSpeed|) at a given temperature. For full listing of functions see Table~ \ref{tab:alkaliatom} and full documentation in Supplemental~\cite{arcDocumentation2016}.

\subsection[Single-atom calculations]{Single-atom calculations}

This module \texttt{calculations\_atom\_single} provides calculations of single-atom properties. It supports plotting of atom energy levels with interactive finding of associated transition wavelengths and frequencies (Sec.~\ref{sec:levelpot}), i.e. Grotrian diagrams, and atom-energy-level shifts in electric fields (Sec.~\ref{sec:stark}), i.e. Stark maps.

\subsubsection{Level plots}\label{sec:levelpot}

The \verb|LevelPlot| class facilities simple plotting of atomic energy levels. It provides an interactive plot for exploring transition wavelengths and frequencies. For example, generation and plotting of the ceasium energy level diagram, including $\ell$ states from $S$ to $D$ for principal quantum numbers from $n=6$ to $n=60$, can be realized with:
\begin{lstlisting}
atom = Caesium()
levels = LevelPlot(atom)
# parameters: nmin, nmax, lmin, lmax
levels.makeLevels(6,60,0,3)
levels.drawLevels()
levels.showPlot()
\end{lstlisting}

This simple example also demonstrates that the more complicated calculations are implemented as a compact classes, whose initialization and methods closely follow naming and stages one would perform in manual calculation. Yet, working on high abstraction level, one can obtain information directly relevant for research in just a couple of high-level commands~\cite{Sibalic2016c}.

\subsubsection{Stark shifts}\label{sec:stark}
Calculation of atomic Stark shifts in external static electric fields provides a tool for both precision electrometry and a mechanism for tuning interatomic interactions to a F\"{o}rster resonance to exploit strong resonant dipole-dipole interactions~(Sec.~\ref{sec:forster}). To find the energy of the atom in an electric field $\mathcal{E}$ applied along the $z$-axis it is necessary to find the eigenvalues of the system described by the Stark Hamiltonian
\begin{equation}\label{eq:starkHamiltonian}
\mathcal{H} = \mathcal{H}_0 + \mathcal{E}\hat{z},
\end{equation}
where $\mathcal{H}_0$ is the Hamiltonian for the unperturbed atomic energy levels and $\mathcal{E}$ is the applied electric field. The electric field term causes a mixing of the bare atomic energy levels due to coupling by the Stark interaction matrix elements $\langle n,\ell,j,m_j\vert \mathcal{E}\hat{z}\vert n',\ell',j',m_j'\rangle$ which can be evaluated from Eq.~(\ref{eq:sphj}) with $q=0$. The selection rules of the Stark Hamiltonian give $\Delta m_j =0,\Delta\ell=\pm1$ such that only states with projection $m_j$ are coupled together, so each Stark map can be constructed by taking basis states with the same $m_j$ value. Following the method of Zimmerman \emph{et al.} \cite{zimmerman79}, Stark shifts are calculated by exact diagonalisation of the Hamiltonian.

\begin{figure}
\includegraphics[width=\columnwidth]{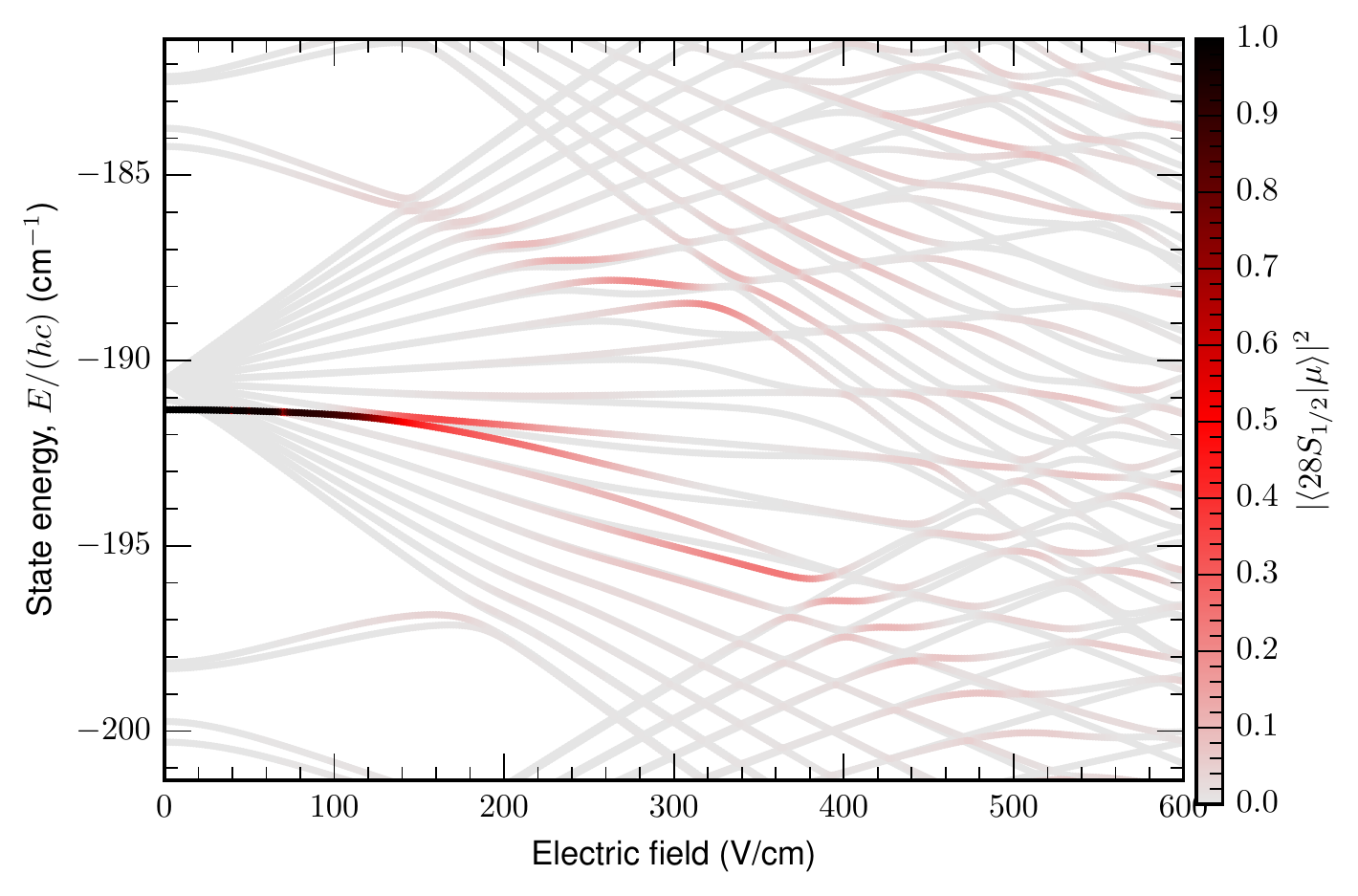}
\caption{\label{fig:starkmap}Example of Stark map calculation, showing caesium $28~S_{1/2}~m_j=1/2$ state perturbation by DC electric field. Colour highlights contribution of the $|28~S_{1/2}~m_j=1/2\rangle$ state in the atom eigenstates $|\mu\rangle$. \label{fig:stark}}
\end{figure}

Stark shift calculations are handled using the \verb|StarkMap| class, which is initialised by passing the appropriate \verb|atom| class. For a target state $\vert n,\ell,j,m_j\rangle$, a range of $n$ values from $n_{\rm min}$ to $n_{\rm max}$ with values of $\ell$ up to $\ell_{\rm max}$ is required to define the basis states for the Stark Hamiltonian. Convergence is typically achieved using $\ell_{\rm max}$ of 20 and $n_{\rm max}-n_{\rm min}\sim10$, however for large applied fields or higher values of $n$ it will be necessary to increase the basis size to account for the strong mixing of the energy levels. Finally, the Hamiltonian is diagonalised for each value of the electric field (defined in V/cm). As an example, to calculate the Stark map shown in Fig.~\ref{fig:stark} for the $28~S_{1/2}~m_j = 1/2$ state in Cs we include states $n_{\rm min}=23$ to $n_{\rm max} = 32$ with  $l_{\rm max} = 20$ for 600 equidistant electric field values in the range from 0 to 600~V/cm. The corresponding program call is:

\begin{lstlisting}
calc = StarkMap(Caesium())
# parameters: n, l, j, mj, nmin, nmax, lmax
calc.defineBasis(28, 0, 0.5, 0.5, 23, 32, 20)
calc.diagonalise(np.linspace(0.,600*1e2,600))
calc.plotLevelDiagram()
calc.showPlot()
\end{lstlisting}

\noindent In interactive mode, the plot can be clicked to obtain the dominant contribution of the basis states within each eigenstate and by default the figure is highlighted proportionaly to the fractional contribution of the target state (in this case $28~S_{1/2}~m_j=1/2$). An alternative option is to highlight the plot proportional to the transition probability for laser excitation from state \linebreak $\vert n',\ell',j',m_{j'}\rangle$ with a laser polarisation driving $q$ polarised transition (where $q=\pm 1,0$ corresponds to $\sigma^\pm$, $\pi$). This is enabled using the optional parameter \lstinline|drivingFromState| in the call to the \verb|diagonalise| method of \verb|StarkMap|. For example, the corresponding code for driving $\sigma^+$ transition from $7~S_{1/2}~m_j=-1/2$ would be

\begin{lstlisting}
 calc.diagonalise(np.linspace(00.,6000,600), drivingFromState=[7,0,0.5,-0.5,+1])
\end{lstlisting}

Finally, to evaluate the static polarizability $\alpha_0$ of the target state in units of MHz/(V/cm)$^2$, the  \verb|getPolarizability| method is called on the \verb|StarkMap| object, i.e.

\begin{lstlisting}
 print("%.5f MHz cm^2 / V^2 " % calc.getPolarizability())
\end{lstlisting}

\subsection[Pair-state calculations]{Pair-state calculations: }\label{sec:pairstate}

The module \texttt{calculations\_atom\_pairstate} contains classes and methods for the calculation and visualisation of long-range and short range interactions (\verb|PairStateInteractions|), as well as an automated tool for identifiying F\"orster resonances \linebreak(\verb|StarkMapResonances|) for electric field tuning of the interaction potential.

\subsubsection{Interatomic interactions}\label{sec:interactions}

Pair-wise interactions between two atoms with internuclear separation $R$, and electron coordinates $\mathbf{r}_1$ and $\mathbf{r}_2$ relative to the respective nuclei, can be expanded in multipole form as \cite{singer05}
\begin{equation}
V(R) = \displaystyle\sum_{L_1,L_2=1}^\infty \frac{V_{L_1, L_2}(\mathbf{r}_1, \mathbf{r}_2)}{R^{L_1+L_2+1}},
\end{equation}
where $L_1+L_2=2$, $L_1+L_2=3$ and $L_1+L_2=4$ correspond respectively to dipole-dipole, dipole-quadrupole and quadrupole-quadrupole interactions, and
\begin{equation}\label{eq:vl1l2}
\begin{split}
V_{L_1, L_2}(\mathbf{r}_1,\mathbf{r}_2) = &\frac{(-1)^{L_2} 4\pi}{\sqrt{(2 L_1+1)(2 L_2+1)}}\\
\times&\displaystyle\sum_m\sqrt{\begin{pmatrix}L_1+L_2\\ L_1+m\end{pmatrix}\begin{pmatrix}L_1+L_2\\L_2+m\end{pmatrix}}r_1^{L_1} r_2^{L_2} Y_{L_1,m}(\hat{r}_1)Y_{L_2,-m}(\hat{r}_2),
\end{split}
\end{equation}

\noindent where $(\ldots)$ is the binomial coefficient and $Y_{L,m}(\hat{r})$ spherical harmonics. In the Eq.~(\ref{eq:vl1l2}) quantization axis is oriented along internuclear axis $\mathbf{R}$. 
On the other hand, atomic states' quantization axis is typically defined with respect to the driving laser, being directed along the laser propagation direction for circularly polarized laser beams, or in plane of electric field vector, perpendicular to the laser propagation direction, for the linearly polarized driving, or in direction of applied static electric and magnetic fields. When the atomic quantization axis is along the internuclear axis, matrix elements $\langle j_1m_{j1}|V_{L_1,L_2}|j_2, m_{j2}\rangle$ are easily evaluated.
When quantization axis is not oriented along the $\mathbf{R}$, their relative orientation can be described  with polar angle $\theta$ and azimuthal angle $\phi$ that the inter-atomic axis makes with quantization axis [Fig.~\ref{fig:orientation_anisotropy}(a)]. Atom states $|j,m_j\rangle$ are then rotated with WignerD matrices $w_{\rm D}(\theta,\phi)$ (\verb|wignerDmatrix|), so that inter-atomic axis defines quantisation direction~\cite{sobelman79}. The coupling can then be easily calculated between the rotated states $|s\rangle = w_{\rm D}(\theta,\phi) |j,m_j\rangle$. Note that $|j,m_j\rangle$ in rotated basis $|j',m_j'\rangle$ is now, in general case, superposition of $m_j'$ states. 

This multipole expansion is valid as long as the wavefunctions of the atoms do not overlap, which is the case for interatomic separations $R$ larger than the Le Roy radius \cite{leroy74}
\begin{equation}\label{eq:leroy}
R_\mr{LR}=2\left(\langle r^2_1 \rangle^{1/2}+\langle r^2_2 \rangle^{1/2}\right)
\end{equation}

\noindent Evaluation of the Le Roy radius can be achieved using the function \texttt{getLeRoyRadius}. For example Le Roy radius for two Ceasium atoms in $n~S$ state is $0.1~\mu$m for $n \sim 20$ and reaches $1~\mu$m for $n \sim 60$, marking interatomic distance below which the results of the pair-state diagonalisation become invalid.

To understand the effect of interactions, consider a pair of atoms in Rydberg pair-state $\vert r, r\rangle$ coupled via $V(R)$ to Rydberg pair-state $\vert r',r''\rangle$ which has an energy defect $\Delta_{r',r''} =2E_{r}-E_{r'}-E_{r''}$.  In the pair-state basis $\{|rr\rangle, |r'r''\rangle\}$ their interaction is described with the hamiltonain

\begin{equation}
\mathcal{H}_{\rm int} = \begin{pmatrix}
0 & V(R)\\
V(R) & \Delta_{r'r''}
\end{pmatrix}.
\end{equation}

\noindent We see that at short range where $V(R)\gg\Delta_{r',r''} $ the splitting of the energy eigen-states $\pm V(R)$
 is dominated by the pair-state interaction energy.  Assuming that V(R) has non-zero dipole-dipole term of multipole expansion [$L_1+L_2=2$ in Eq.~(\ref{eq:vl1l2})], this corresponds to resonant dipole-dipole regime, giving eigenstates' energy distance dependance $\propto C_3/R^3$. At large $R$ where $\Delta_{r',r''}\gg V(R)$, the interaction is second order leading to an energy shift of $-V(R)^2/\Delta_{r',r''}=-C_6/R^6$, known as the van der Waals regime. The cross-over between these regimes occurs at the van der Waals radius $R_{\rm vdw}$ where $V(R_{\rm vdw})=\Delta_{r',r''}$.

For a real system, whilst the pair-state with the smallest $\Delta_{r',r''}$ dominates the resulting interaction shift, it is necessary to consider the effects of all near-resonant pair-states that are coupled by the $V(R)$ interaction term above. Calculations of these pair-wise interaction potentials are handled by the \linebreak\verb|PairStateInteractions| class that is initialised by specifying the element name and target pair-state $n_1,l_1,j_1, n_2, l_2, j_2, m_{j_1}, m_{j_2}$ whose behaviour we want to explore.

\begin{lstlisting}
 calc=PairStateInteractions(atom(),n1,l1,j1,n2,l2,j2,mj1,mj2)
\end{lstlisting}

\subsubsection{Long-range limit}

At large separation, for off-resonantly ($\Delta_{r',r''}\neq0$) coupled states, the dipole-dipole interactions dominate to give an interaction of the form $-C_6/R^6$ van der Waals potential where the sign depends on the energy defect of the closest dipole-coupled pair-states $|r'r''\rangle$, leading to attractive or repulsive interactions accordingly. In the long-range limit $V(R)\ll \Delta$ (where $\Delta$ is the energy defect of the closest dipole-coupled pair-state), the $C_6$ coefficient for pair-state $|rr\rangle$ can be evaluated using second-order perturbation theory as
\begin{equation}
C_6 = \sum_{r',r''}\frac{|\langle r' r'' | V(R) |rr\rangle|^2}{\Delta_{r',r''}},
\end{equation}

\begin{figure}[b!]
\includegraphics[width=\columnwidth]{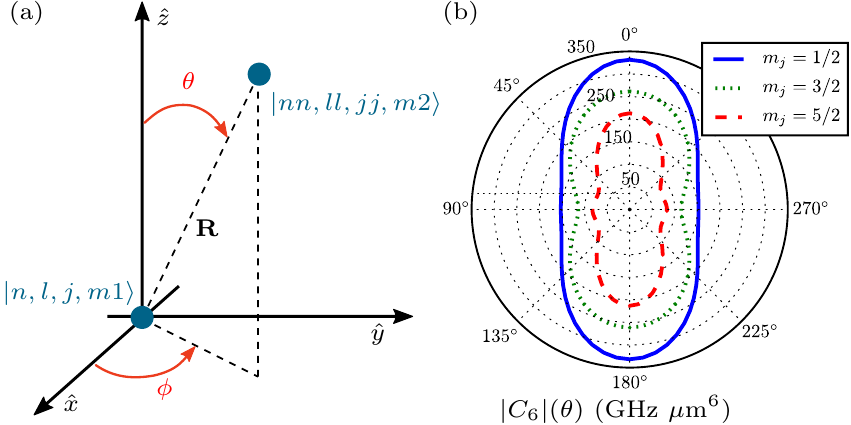}
\caption{\label{fig:orientation_anisotropy} Orientation of the two interacting atoms and their anisotropic interactions. (a) Orientation of the two atoms is defined as a polar $\theta$ and axial $\psi$ angle that inter-atomic axis makes with respect to the quantization axis $\hat{z}$. (b) Anisotropy of long-range interactions is here illustrated in the case of two rubidium atoms, both in $60~D_{5/2}~m_j$ state, through the dependance of $C_6$ interaction coefficient on angle $\theta$ between quantisation axis and inter-molecular axis.}
\end{figure}

\noindent where the sum runs over all pair-states $|r'r''\rangle$ whose energy differs from the pair-states $|rr\rangle$ energy for $\Delta_{\rm r',r''}<\Delta_{\rm E}$, where $\Delta_{\rm E}$ is some maximal energy defect that provides a truncation of the basis states. This calculation is performed using the method \linebreak \verb|getC6perturbatively|, returning $C_6$ in units of GHz~$\mu$m$^6$. Users have just to specify the relative orientation of the atoms, the range of principal quantum numbers for the states used in calculation, and the maximal energy defect. For example for the interaction of two rubidium atoms in $60~S_{1/2}$, $m_{j_1}=1/2$, $m_{j_2}=-1/2$ states, whose inter-atomic axis is set at an angle of $\theta=\pi/6$ with respect to the quantisation axis [Fig.~\ref{fig:orientation_anisotropy}(a)], the $C_6$ interaction term can be calculated as perturbation of states with principal quantum number differing maximally $\delta n = 5$ from the $n=60$, and maximal energy difference in pair-state energies of $h\times 25$~GHz as
\begin{lstlisting}
# parameters: atom, n1, l1, j1, n2, l2, j2, mj1, mj2
calc =  PairStateInteractions(Rubidium(), 60, 0, 0.5, 60, 0, 0.5, 0.5, -0.5)
# parameters: theta, phi, deltan, deltaE
c6 = calc.getC6perturbatively(pi/6, 0, 5, 25.e9);
print("C_6  = %.0f GHz (mu m)^6" % c6)
\end{lstlisting}

Using this function, the anisotropy of the $V(R)$ interaction can be easily identified as shown in Fig.~\ref{fig:orientation_anisotropy}(b) which plots the magnitude of $C_6$ for a pair of atoms in the $60~D_{5/2}$ state of Rubidium for $\theta=0\ldots2\pi$.

\subsubsection{Exact interaction potential}

To evaluate the interaction potential for arbitrary separation [valid down to the Le Roy radius, Eq.~(\ref{eq:leroy})], it is necessary to diagonalise the matrix $V(R)$ containing all the interatomic couplings for each separation $R$ to obtain exact values of the eigenvalues and eigenstates describing the interaction between atomic pair-states. Due to the strong admixing of states, this process requires careful choice of atomic basis states, including higher order orbital angular momentum states up to $\ell_{\rm max}$. For each distance $R$, the matrix is diagonalised using an efficient ARPACK package provided through Numpy, and the $n_{\rm eig}$ eigenstates whose eigenvalues are closest to the target pair-state are returned.

\begin{figure}
\includegraphics[width=\columnwidth]{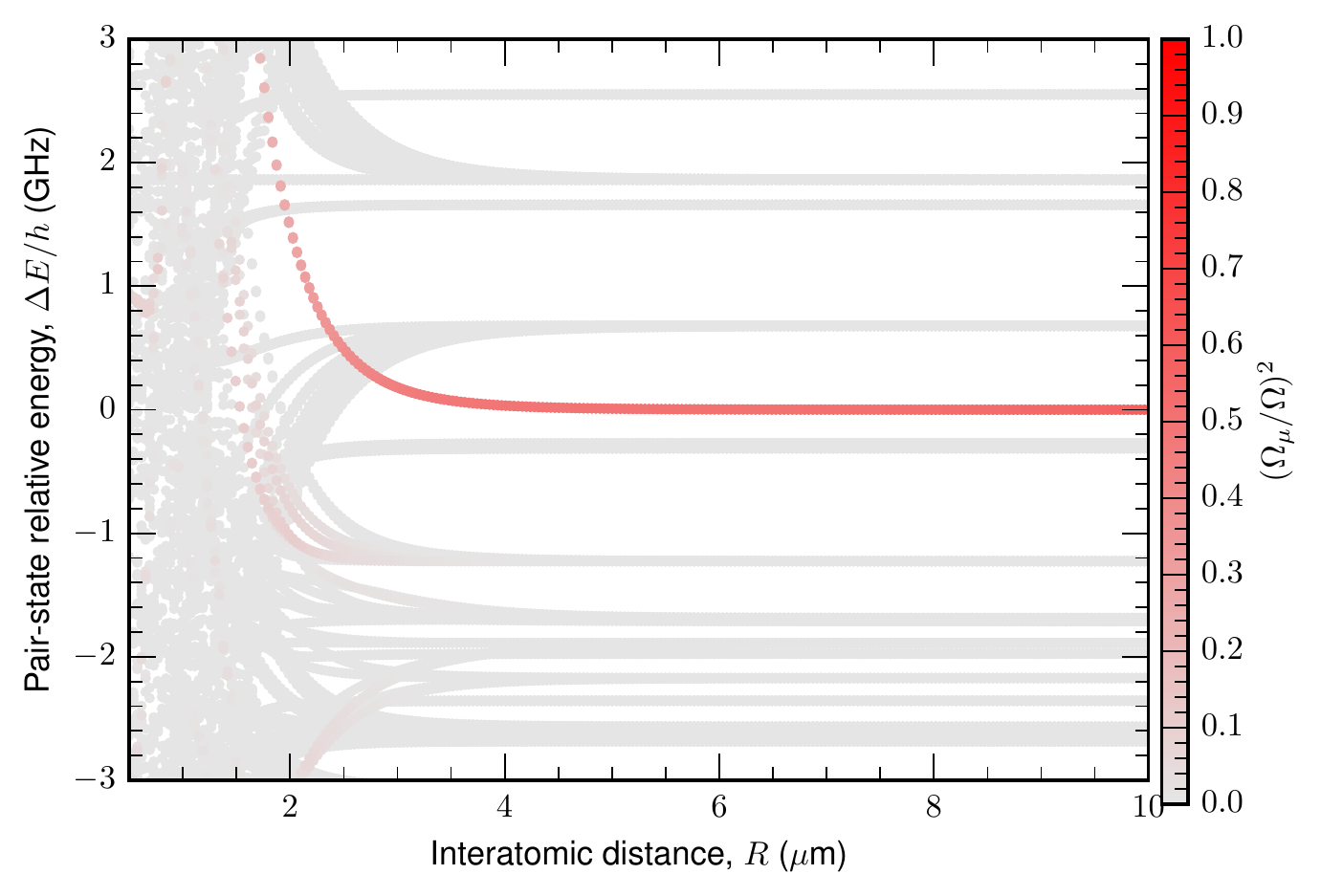}
\caption{\label{fig:interactions}Example of interactions between the atoms, causing level shifts in of the atomic pair-states. Here shown exemplary in vicinity of Rubidium $|60~S_{1/2}~m_j=1/2,60~S_{1/2}~m_j=-1/2$ state. Highlighting is proportional to driving strength from $5~P_{3/2}~m_j = 3/2$ state with coupling laser driving $\sigma^-$ transitions. Atom interatomic direction is oriented along the quantization axis ($\theta,\phi = 0$). Using provided methods we can find van der Waals radius $R_{\rm vdW} = 2.4~\mu$m, and short-range $C_3= 16.8~$GHz$~\mu$m$^3$ and long range $C_6=135~$GHz$~\mu$m$^6$ dispersion coefficients.}
\end{figure}

Figure.~\ref{fig:interactions} provides an example of the interaction potential for a pair of atoms initially in the $|60~S_{1/2}~m_j=1/2,60~S_{1/2}~m_j=-1/2\rangle$ state of Rubidium for 200 interatomic spacings in the range of 0.5-10~$\mu$m, finding $n_{\rm eig}=150$ closest eigenstates, accounting for states with orbital angular momentum up to $l_{\rm max} = 4$ created using the code

\begin{lstlisting}
calc =  PairStateInteractions(Rubidium(), 60, 0, 0.5, 60, 0, 0.5, 0.5, -0.5)
# parameters: theta, phi, nMax, lMax, maxEnergy
calc.defineBasis (0, 0, 5, 4, 25.e9)
calc.diagonalise (np.linspace (0.5,10.0,200) ,150)
calc.plotLevelDiagram()
calc.showPlot()
\end{lstlisting}

As with the \verb|StarkMap| method, the figure is interactive allowing users to determine the dominant composition (expressed in the pair-state basis) of a given eigenstate. The default behaviour is to highlight the eigenvalues proportional to the fractional contribution of the original target pair-state. Alternatively, as shown on Fig.~(\ref{fig:interactions})], the optional parameter \linebreak\verb|drivingFromState| is used in the call to \verb|diagnolise| to give highlighting proportionaly to the relative laser coupling strength from a given state, assuming that one atom is already in one of the two states that make the initially specified pair-state.

By default, \verb|PairStateInteractions| includes only dipole coupling between the states in calculating level diagrams. Interactions up to quadrupole-quadrupole term [including all terms $L_1+L_2 \leq 4$ in Eq.~(\ref{eq:vl1l2})] can be included by setting optional parameter \verb|interactionsUpTo = 2| during the \linebreak \verb|PairStateIntearctions| initialization, which can be important for short-distance structure of level diagram \cite{Deiglmayr2014}. For evaluation of long-range potential curves this additional term makes only small perturbations to the asymptotic $C_6$ behaviour. However, for accurate determination of the molecular levels of the short-range potential wells it is important to include the higher order multipole terms. Convergence should also be checked by increasing the basis size and re-evaluating the relevant parameters to ensure all the relevant states are included in the calculation basis.

Following diagonalisation of the pair-state interaction matrix, the long-range ($C_6$) and short range ($C_3$) dispersion coefficients can be evaluated using the methods \linebreak\verb|getC3fromLevelDiagram| and \verb|getC6fromLevelDiagram| respectively, which perform a fit to the eigen-energies associated with the state containing the largest admixture of the target state. A method for finding the cross-over distance between van der Waals and resonant dipole-dipole interactions, i.e. van der Waals radius $R_{\rm vdw}$, is also provided (\verb|getVdwFromLevelDiagram|).

\subsubsection{Stark tuned F\"orster resonances}\label{sec:forster}

\begin{figure}[b!]
\includegraphics[width=\columnwidth]{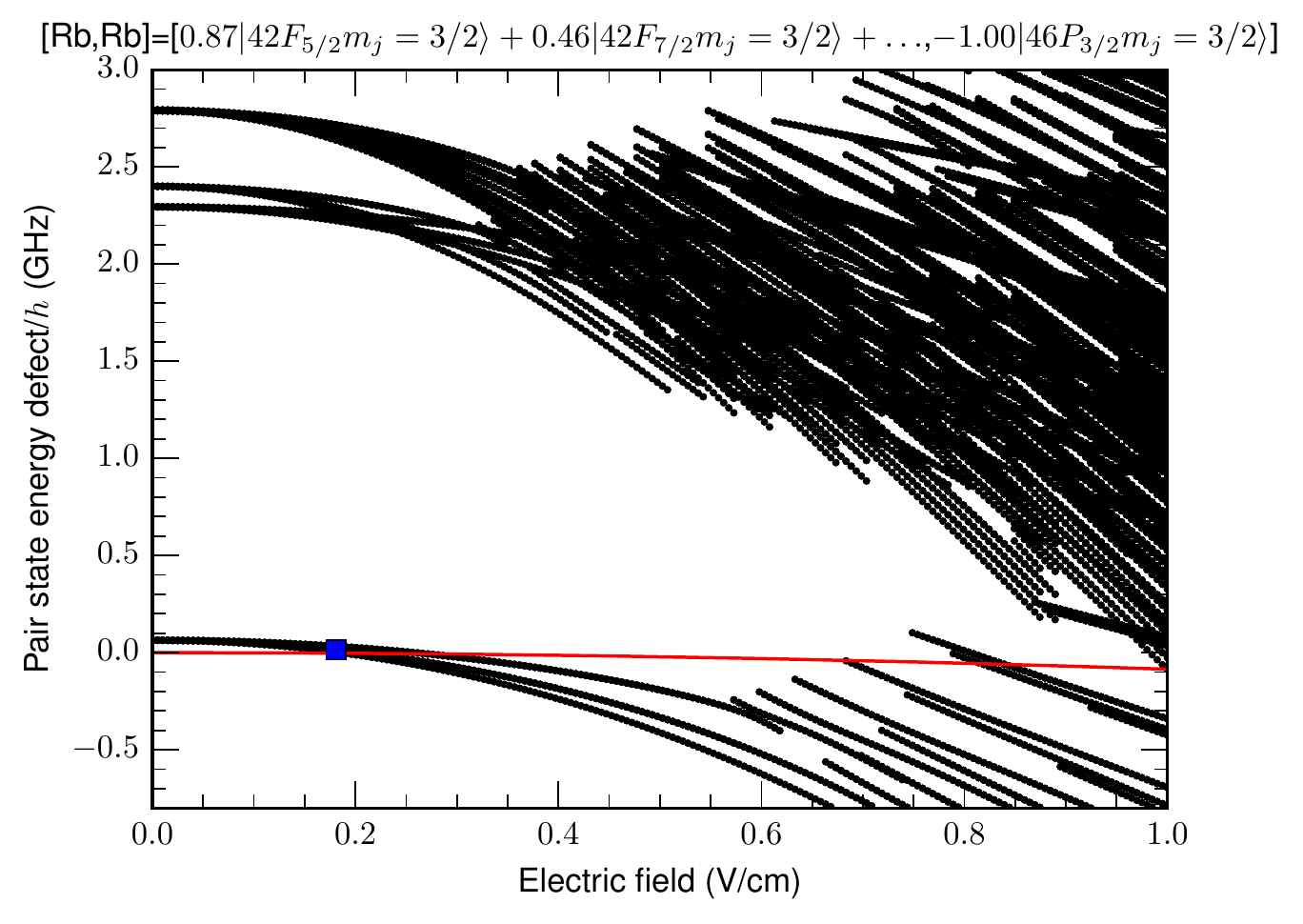}
\caption{\label{fig:forster}(color online) Example of plot produced by \texttt{StarkMapResonances} for finding F\"orster resonances for Rubidium pair-state $44~D_{5/2}~m_j=5/2$ shown as grey (red) line. Pair-states whose main admixed state is dipole-coupled to the original pair-state are shown as black lines. Clicking on one of this states, marked with square (blue) on the plot, reveals it's composition in the plot title. In this case, it's a mixture of $42~F_{5/2}$ and $42~F_{7/2}$ states in one of the atoms, and essentially unperturbed $46~P_{3/2}$ state that is almost resonant with original pair-state at electric field of $E\approx 0.18$~V/cm.}
\end{figure}

As outlined in Sec.~\ref{sec:interactions}, the finite pair-state energy defects $\Delta$ associated with the closest dipole-coupled pair-states lead to a transition from first order resonant dipole-dipole interactions ($\propto R^{-3}$) to second order van der Waals ($\propto R^{-6}$) at the van der Waals radius. Using external electric fields however, it is possible to Stark shift the pair-states into resonance to obtain long-range $\propto R^{-3}$ interactions for all values of $R$, known as a F\"orster resonance \cite{gallagher82,vogt06,ravets15,Beterov2015}.

To identify suitable F\"orster resonances the \linebreak \verb|StarkMapResonances| class is used, taking in a pair of target pair-states and performing diagonalisation of the Stark Hamiltonian of Eq.~(\ref{eq:starkHamiltonian}) in the pair-state basis. Due to the angular momentum selection rules of the dipole operator $V(R)$, the target pair-state can be dipole-coupled to pair-states with $\Delta m_j=\pm1,0$, it is necessary to calculate Stark maps for a range of $m_j$ manifolds. Following diagonalisation, only pair-states with energy close to the target state are considered, with any states not dipole-coupled to the target pair-state being discarded. As the electric field coupling leads to significant mixing of the zero-field pair-states, the algorithm identifies the basis state containing the largest target state fraction at a given electric field to test for the closest dipole-coupled pair-states. Finally, an interactive plot routine enables users to identify states that have a F\"orster resonance. Note, unlike the previous class, \linebreak \verb|StarkMapResonances| accepts two atom classes at initialisation making it possible to determine inter-species resonances.

Fig.~\ref{fig:forster} shows an example plot to identify F\"orster resonances for a pair of atoms in the $44~D_{5/2} m_j=5/2$ state of Rubidium that is generated using the following code

\begin{lstlisting}
state = [44,2,2.5,2.5]
calculation = StarkMapResonances(Rubidium(), state, Rubidium(), state)
# nMin, nMax, lMax, rangeEfield, [energyRange]
calculation.findResonances(39, 49, 20, np.linspace(0,100,200), energyRange=[-0.8e9,3.e9])
calculation.showPlot()
\end{lstlisting}

\noindent In the above example, similarly as when finding a Stark map (Sec.~\ref{sec:stark}), we have to specify a range of acceptable principal quantum numbers $[n_{\rm min},n_{\rm max}]$ and maximal orbital angular momentum $l_{\rm max}$ for the basis states for Stark map calculations, as well as the electric field range (in this case 0-1~V/cm at 200 equidistant points). An additional argument \verb|energyRange| defines the energy window within which we will keep the resonant pair-states. In the above given example, this is in range $h \times [-0.8,3.0]$~GHz. From the interactive plot, selection of the pair-state eigenvalues shows a F\"orster resonance occuring at 0.18~V/cm with the $42~F,46~P_{3/2}$ pair-state where due to the electric coupling the F state is an admixture of the $42~F_{5/2,7/2}~m_j=3/2$ states.

\section{Installation and Usage}\label{sec:getting_started}

It is assumed the the pre-requisites (Numpy, SciPy and Matplotlib) are installed and can be located by Python interpreter (see e.g. \cite{anaconda}).
Both Python 2.7 and 3.5 are supported. To achieve good performance, it is recommended to use Numpy packages that connect to optimised backends, like ATLAS \cite{Whaley1998}. Prepackaged Python distributions, like Anaconda \cite{anaconda}, provide this out-of-the box. The ARC library can be downloaded online \cite{arcGitHub} as a .zip file release. Installation is performed by extracting the downloaded .zip archive and copying the arc subfolder into the root of your project directory. It is important that Python has write access to the folder where the package is located, so that database files (stored in \verb|arc/data/|) can be updated and used. 

\subsection{Optimised Numerov integrator}
Integration of the atomic wavefunctions to obtain dipole matrix elements is numerically intensive and by default ARC uses an optimised Numerov integration routine implemented as C extensions of Python (\verb|arc_c_extensions|). In the unlikely case the precompiled executable \texttt{arc\_c\_extensions.so} is incompatible with the installed system, please install a C compiler and compile \texttt{arc\_c\_extensions.c} located in the ARC root folder using by calling from command line
\begin{lstlisting}
python setupc.py build_ext --inplace
\end{lstlisting}
Note that path to arc directory should not contain spaces in order to \texttt{setupc.py} script to work. Alternatively, the native Python solver can be used
by setting the optional argument \texttt{cpp\_numerov=False} when initialising the atom class, however this is not recommended for intensive calculations as it results in substantially slower execution.

\subsection{Getting started}

To initialise the library, use the following code at the start of a Python script or interactive IPython notebook with
\begin{lstlisting}
# locate ARC Directory
import sys, os
rootDir = '/path/to/root/directory/for/arc'
sys.path.append(rootDir)
os.chdir(rootDir)
# import ARC library
from arc import *
\end{lstlisting}

\noindent This firsts sets a path to the directory containing ARC package directory on your computer\footnote{Note that in the directory path, a backslash (\textbackslash) should be used on Windows machines, instead of forward slash used on UNIX based machines (Linux, MacOS).}.  This is recommended way of using the package in research environment, since users can easily access, check and change the underlying code and constants according to their needs. ARC is now ready for use, and can be tested using the example code above. Numerous additional examples are provided in IPython notebook "An Introduction to Rydberg atoms with ARC"~\cite{Sibalic2016c}.

\section{Implementation}\label{sec:implementation}

\subsection{Physical constants}

As mentioned above, the atomic properties are encapsulated in classes which contain the relevant atomic properties determined from the literature along with model potential coefficients taken from Marinescu et al. \cite{Marinescu1994} which have been optimised against measured energy levels. For each atom, asymptotic expansions of the quantum defects are used to determine the energy levels of states with high principal quantum number to high accuracy, using measured values the ionisation energy, Rydberg constant and quantum defects taken from Li~\cite{Goy1986,Lorenzen1983}, Na~\cite{Lorenzen1983}, K~\cite{Lorenzen1983}, Rb \cite{li03,han06,Mack2011,Afrousheh2006a}, and Cs \cite{goy82,Weber1987,deiglmayr16}. For the low-lying states, the quantum defects do not accurately reproduce the measured energies and instead energy levels are determined from data in the NIST ASD database \cite{NISTasd_ver5_4}. By default, the cut-off between tabulated and calculated energies is determined by the point at which the error in the calculated energy exceeds 0.02\%.

\subsection{Tracking calculation progress}
For tracking progress of calculations, most functions (see documentation~\cite{arcDocumentation2016} for details) provide optional \verb|bool| arguments, that turn on printing of additional information about calculations. For example, \verb|progressOutput| prints basic status information, and is recommended for everyday use. If \verb|debugOutput| is set to \verb|True|, more verbose information about basis states and couplings will be printed in the standard output.

\subsection{Memoization}
In order to achieve good performance, memoization is used throughout. That is, when functions receive request for calculation, memorised previous results are checked, and the value is retrieved from memory if it already exists. Both in-application SQL database (SQLite) and standard arrays are used for this. Memoization is used for dipole and quadrupole matrix element calculations and all sorts of angular coupling factors, that are independent of principal quantum number of the considered state, including Wigner-$n$J coefficients and WignerD matrices. Single-atom and pair-state calculations automatically will update databases if new values are encountered, speeding up the future calculations. Updating dipole and quadrupole matrix element database can be done manually too, by calling \linebreak\verb|updateDipoleMatrixElementsFile|.

\subsection{Saving and retrieving calculations}
Large self-contained calculations, such as Stark maps or pair-state interaction potentials, are wrapped as classes to enable easy access to the calculations parameters, with associated visualisation and exploration methods to make the results easy to disseminate. An added benefit is that calculations can be saved for future use, as illustrated in the example below:

\begin{lstlisting}
calc = PairStateInteractions(Rubidium(),
                          60,0,0.5,60,0,0.5, 0.5,0.5)
# do something with initialized calculation
calc.defineBasis(0,0, 5,5, 25.e9)
calc.diagonalise(np.linspace(0.5,10.0,200),150)
|\textbf{saveCalculation}|(calc, "myCalculation.pkl")
\end{lstlisting}

\noindent This calculation can then be retrieved at a later time, or even on another machine, and continued, or further explored. For example:

\begin{lstlisting}
calc = |\textbf{loadSavedCalculation}|("myCalculation.pkl")
# continue calculation
calc.plotLevelDiagram()
calc.showPlot()
\end{lstlisting}

\subsection{Exporting data to file}

If one wants to use the obtained results in another program, for further analysis and calculation, both \verb|PairStateInteractions| and \verb|StarkMap| provide \verb|exportData| method that allows saving data in .csv format, that is human-readable and easy to import in other software tools. As a commented header of the exported files, program will record details of the performed calculations, in human-readable form. Provided that initial calculation is variable \verb|calc|, its data is exported by calling

\begin{lstlisting}
calc.exportData("rootFileName")
\end{lstlisting}

\noindent The program will output list of files (typically three files) storing the calculation data, whose names will be starting with\linebreak \verb|"rootFileName"|.

\subsection{Advanced interfacing of ARC with other projects}

In addition to graphical exploration, and exporting of relevant data as .csv files, advanced users incorporating ARC in their own projects might want to access directly Stark or pair-state interaction matrices, and corresponding basis states that are generated by \verb|defineBasis| methods of the corresponding calculation classes. Basis (pair-) states are accessible for both methods in \verb|basisStates| array. Stark matrix can be assembled as sum of diagonal matrix \verb|mat1| recording state energies, and off-diagonal matrix that depends on applied field and can be obtained as product of electric field and \verb|mat2|. For pair-state interactions, the interaction matrix can be obtained as a sum of interatomic-distance independent matrix \verb|matDiagonal|, recording energy defects of pair-states, and distance dependant matrices stored in \verb|matR|, where \verb|matR[0]|, \verb|matR[1]| and \verb|matR[2]| store respectively dipole-dipole, dipole-quadrupole and quadrupole-quadrupole interaction coefficients for interaction terms [Eq.~(\ref{eq:vl1l2})] scaling respectively with distance $R$ as $\propto R^{-3}$,  $\propto R^{-4}$ and $\propto R^{-5}$. For more details about this, examples of use, and other accessible calculated information, we refer the reader to the detailed documentation~\cite{arcDocumentation2016}.

\section{Outlook}\label{sec:outlook}

This paper describes version 1.2 of ARC that can be downloaded from~\cite{atomcalcWebPage}. Supplementary information provides full documentation~\cite{arcDocumentation2016} for functions listed in \ref{app:function_listing}, as well as IPython notebook~\cite{Sibalic2016c} that contains more elaborate examples, code snippets and benchmarks against published results. This notebook provides both good introduction to most of the capabilities of the ARC, as well as good starting point for users, providing useful code snippets. Numerous examples also provide quantitative interactive introduction for anyone starting in the field of Rydberg atomic physics. In addition, for quick estimates, when one is in the lab, conference or other meeting, we provide web-interface to the package~\cite{atomcalcWebPage}. For latest versions of the package and documentation, please download material from the ARC GitHub page~\cite{arcGitHub}.

In the future, the package can be extended to include calculations of dressing potentials~\cite{Bijnen2015}, magic wavelengths~\cite{Goldschmidt2015a}, atom-wall interactions~\cite{Derevianko1999,Bloch2005}, photoionisation, collisional cross-sections~\cite{Kiffner2015}, tensor polarizability, molecular bound states~\cite{Gaj2014}, effects of magnetic fields, microwave tuning of interactions~\cite{Sevincli2014} and other atomic properties. These can be added as calculation tools, in separate classes, built on top of the existing library. Alkaline earth elements can be included too. While these were beyond the scope of the current project, we hope that this project can provide an initial seed for a much bigger community project. The code is hosted on GitHub~\cite{arcGitHub}, allowing easy community involvement and improvements. A proposal of some basic philosophy for the development is provided in documentation~\cite{arcDocumentation2016}.

We hope the library will increase accessibility to the existing knowledge. Up to now a lot of relevant information, although in principle derivable from existing literature, required quite lengthy and error-prone calculations. The developed hierarchical object-oriented structure allows one to retrieve relevant information at the appropriate level of abstraction, without the need to deal with lower-level details. In addition, we hope to establish growing code base for common calculations, in the spirit of other open-source community projects like, Numpy and SciPy~\cite{Oliphant2007}, adding to the growing stack of atomic physics tools like ElecSus~\cite{Zentile2015}, The Software Atom~\cite{Javanainen2016} and QuTIP~\cite{Johansson2013}. We especially highlight recent related development of C++ program Pairinteraction for pair-state calculations~\cite{swpi2016}. We hope that this efforts will allow more research groups to explore the rich physics achievable when one uses all the available transitions in the atoms. In addition to exploring Rydberg physics, library can be useful for practical exploration of multi-photon schemes for alkali-atoms excitation~\cite{Moon2007,Lee2013b,Kondo2015}, ionization~\cite{Courtade2004,Ates2013} and upconversion~\cite{Akulshin2009,Sulham2010,Akulshin2014}, as well as state control, e.g. through dressing of the intermediate levels~\cite{Sibalic2016b} and adiabatic transfers~\cite{Maas1999}.

\section*{Acknowledgements}

We thank D.~Whiting, J.~Keaveney, H.~Busche, P.~Huillery, T.~Billam, C.~Nicholas, I.~Hughes, R.~Potvliege and M.~Jones for stimulating discussions. This work is supported by Durham University, EPSRC grant EP/N003527/1, FET-PROACT project "RySQ"  (H2020-FETPROACT-2014-640378-RYSQ), EPSRC grant "Rydberg soft matter"  (EP/M014398/1), DSTL and EPSRC grant EP/M013103/1. In the final stages of this project we've been informed by Sebastian Weber about a related efforts on C++ program for pair-state calculations~\cite{swpi2016}.




\appendix

\section{Example program}\label{app:example_code}

Code below generates Fig.~\ref{fig:bbr}. For many more examples see Supplemental~\cite{Sibalic2016c}.

\begin{lstlisting}
from arc import *

# ==== GENERATE DATA (ARC) ====

atom = Rubidium()

pqn = []  # principal quantum number
y = []      # rate at T =0 K
ybb = []  # additional black-body induced transitions

# calculating decay from from 30 S_{1/2}
# to n P_{1/2} and n P_{3/2}
# where n is in range form 5 to 40

for n in xrange(5,40):
    pqn.append(n)

    # transition rate at T = 0K
    noBBR = atom.getTransitionRate(30, 0, 0.5,\
                            n, 1, 0.5, temperature = 0)\
                + atom.getTransitionRate(30, 0, 0.5,\
                            n, 1, 1.5, temperature=0 )

    # same, now at T= 300 K
    withBBR =  atom.getTransitionRate(30, 0, 0.5,\
                            n, 1, 0.5, temperature=300.0)\
                   + atom.getTransitionRate(30, 0, 0.5,\
                            n, 1, 1.5, temperature=300.0)
    y.append(noBBR)
    ybb.append(withBBR-noBBR)

# ==== PLOT(matplotlib) ====

pqn=np.array(pqn)
y = np.array(y)
ybb = np.array(ybb)

width = 0.4
f, ax = plt.subplots()
ax.bar(pqn-width/2.,y,width=width,color="r")
ax.bar(pqn+width/2.,ybb,width=width,color="g")
ax.set_xlabel("Principal quantum number, $n$")
ax.set_ylabel(r"Transition rate (s${}^{-1}$)")
plt.legend(("Spontaneous decays","Black-body induced transitions"),fontsize=10)
plt.xlim(4,40)

f.set_size_inches(5.50,4)
# save figure in decays.pdf
plt.savefig("decays.pdf", bbox_inches='tight')

\end{lstlisting}

\section{ARC function list}\label{app:function_listing}

This appendix provides listings of functions, classes and methods contained within the ARC module. For detailed documentation, and more elaborate examples we refer the reader to the documentation in supplemental~\cite{arcDocumentation2016}, also accessible via ARC website \cite{atomcalcWebPage}.

\begin{table*}[h!]
\caption{\label{tab:alkalifunc}Class and function listing of  \texttt{alkali\_atom\_functions} module}
\centering
\footnotesize
\begin{tabular}{l l}
\hline
Name (parameters) & Short description\\
\hline
\verb|AlkaliAtom|([preferQuantumDefects, cpp\_numerov])  & Implements general calculations for alkali atoms (see Table~\ref{tab:alkaliatom})\\
\verb|NumerovBack|(innerLimit, outerLimit, kfun, ...) & Full Python implementation of Numerov integration\\
\verb|saveCalculation|(calculation, fileName) & Saves calculation for future use\\
\verb|loadSavedCalculation|(fileName) &  	Loads previously saved calculation\\
\verb|printStateString|(n, l, j) &	Returns state spectroscopic label for $|n,l,j\rangle$ \\
\hline
\end{tabular}
\end{table*}

\begin{table*}[ht]
\caption{\label{tab:alkaliatom} Methods and function listing of  \texttt{alkali\_atom\_functions.AlkaliAtom} class. Typical relative uncertanties are obtained from comparison to measured values.}
\centering
\footnotesize
\begin{tabular}{l l c}
\hline
Name (parameters) & Short description (units) & Typical rel. accuracy\\
\hline
\verb|getDipoleMatrixElement|(n1, l1, ...) &	Reduced dipole matrix element ($a_0 e$)& $ \sim 10^{-2}$\\
\verb|getTransitionWavelength|(n1, l1, ...) &	Calculated transition wavelength in vacuum (m)&  $ \sim 10^{-6}$\\
\verb|getTransitionFrequency|(n1, l1, ...) & Calculated transition frequency (Hz)  &  $  \sim 10^{-6}$\\
\verb|getRabiFrequency|(n1, l1, j1, mj1, ...) &Returns a Rabi frequency (angular, i.e. $\Omega = 2\pi\times \nu$) for resonant excitation  & \\
& with a specified laser beam in the center of TEM00 mode (rad~s$^{-1}$)& $ \sim 10^{-2}$\\
\verb|getRabiFrequency2|(n1, l1, j1, mj1, ...) &Returns a Rabi frequency (angular, i.e. $\Omega = 2\pi\times \nu$) for resonant excitation & \\
& with a specified electric field driving amplitude (rad~s$^{-1}$)& $ \sim 10^{-2}$\\
\verb|getStateLifetime|(n, l, j[, ...]) & Returns the lifetime of the state (s)& $ \sim 10^{-2}$\\
\verb|getTransitionRate|(n1, l1, j1, n2, ...) & Transition rate due to coupling to vacuum modes (black body included) (s$^{-1}$)& $ \sim 10^{-2}$\\
\verb|getReducedMatrixElementJ_asymmetric|(n1, ...) & Reduced matrix element in $J$ basis, defined in asymmetric notation ($a_0 e$)& $ \sim 10^{-2}$\\
\verb|getReducedMatrixElementJ|(n1, l1, ...) & Reduced matrix element in $J$ basis, symmetric notation ($a_0 e$)& $ \sim 10^{-2}$\\
\verb|getReducedMatrixElementL|(n1, l1, ...) & Reduced matrix element in $L$ basis, symmetric notation ($a_0 e$)& $ \sim 10^{-2}$\\
\verb|getRadialMatrixElement|(n1, l1, ...) & Radial part of the dipole matrix element ($a_0 e$)& $ \sim 10^{-2}$\\
\verb|getQuadrupoleMatrixElement|(n1, ...) & Radial part of the quadrupole matrix element ($a_0^2 e$)& $ \sim 10^{-2}$\\
\verb|getPressure|(temperature) & Vapour pressure at given temperature (Pa)& $\sim (1-5) \cdot 10^{-2}$\\
\verb|getNumberDensity|(temperature) & Atom number density at given temperature (m$^{-3}$)& $\sim (1-5) \cdot 10^{-2}$\\
\verb|getAverageInteratomicSpacing|(...) & Returns average interatomic spacing in atomic vapour (m)& $\sim (1-5) \cdot 10^{-2}$\\
\verb|corePotential|(l, r) & core potential felt by valence electron (a.u)&\\
\verb|effectiveCharge|(l, r) & effective charge of the core felt by valence electron (a.u)&\\
\verb|potential|(l, s, j, r) & core potential, including spin-orbit interaction (a.u)&\\
\verb|radialWavefunction|(l, s, j, ...) & Radial part of electron wavefunction&\\
\verb|getEnergy|(n, l, j) &  	Energy of the level relative to the ionisation level (eV)& $  \sim 10^{-6}$\\
\verb|getQuantumDefect|(n, l, j) & Quantum defect of the level.&\\
\verb|getC6term|(n, l, j, n1, l1, j1, ...) & C6 interaction term for the given two pair-states ($h~ \times~$Hz~m$^6$)&\\
\verb|getC3term|(n, l, j, n1, l1, j1, ...) & C3 interaction term for the given two pair-states ($h~ \times~$Hz~m$^3$)&\\
\verb|getEnergyDefect|(n, l, j, n1, l1, ...) & Energy defect for the given two pair-states, $E(|rr\rangle) - E(|r'r''\rangle)$ (eV)&\\
\verb|getEnergyDefect2|(n, l, j, nn, ll, ...) & Energy defect for the given two pair-states, $E(|r_1r_2\rangle) - E(|r'r''\rangle)$ (eV)&\\
\verb|updateDipoleMatrixElementsFile|() & Updates the file with pre-calculated dipole matrix elements.&\\
\verb|getRadialCoupling|(n, l, j, n1, l1, j1) & Returns radial part of the coupling between two states (dipole,&\\
& quadrupole) ($a_0 e$ or $a_0^2 e$)& $ \sim 10^{-2}$\\
\verb|getAverageSpeed|(temperature) & Average (mean) speed at a given temperature (m/s)&\\
\verb|getLiteratureDME|(n1, l1, j1, n2, ...) & Returns literature information on requested transition&\\
\hline
\end{tabular}
\end{table*}

\begin{table*}[ht]
\caption{Class listing of  \texttt{alkali\_atom\_data} module. All these classes inherit properties of  \texttt{alkali\_atom\_functions.AlkaliAtom} from Table~\ref{tab:alkaliatom}.}
\centering
\footnotesize
\begin{tabular}{l l}
\hline
Name (parameters) & Short description\\
\hline
\verb|Hydrogen|([preferQuantumDefects, cpp\_numerov]) &	Properties of hydrogen atoms\\
\verb|Lithium6|([preferQuantumDefects, cpp\_numerov]) &	Properties of lithium 6 atoms\\
\verb|Lithium7|([preferQuantumDefects, cpp\_numerov]) &	Properties of lithium 7 atoms\\
\verb|Sodium|([preferQuantumDefects, cpp\_numerov]) &	Properties of sodium 23 atoms\\
\verb|Potassium39|([preferQuantumDefects, cpp\_numerov]) & 	Properties of potassium 39 atoms; alias \verb|Potassium(...)|\\
\verb|Potassium40|([preferQuantumDefects, cpp\_numerov]) & 	Properties of potassium 40 atoms\\
\verb|Potassium41|([preferQuantumDefects, cpp\_numerov]) & 	Properties of potassium 41 atoms\\
\verb|Rubidium85|([preferQuantumDefects, cpp\_numerov]) &	Properties of rubidium 85 atoms; alias \verb|Rubidium(...)|\\
\verb|Rubidium87|([preferQuantumDefects, cpp\_numerov]) &	Properties of rubidium 87 atoms\\
\verb|Caesium|([preferQuantumDefects, cpp\_numerov]) &	Properties of caesium 133 atoms\\
\hline
\end{tabular}
\end{table*}

\begin{table*}[ht]
\caption{Method listing of  \texttt{calculations\_atom\_single.LevelPlot}(atomType) class}
\centering
\footnotesize
\begin{tabular}{l l}
\hline
Name (parameters) & Short description\\
\hline
\verb|makeLevels|(nFrom, nTo, lFrom, lTo) &	Constructs energy level diagram in a given range\\
\verb|drawLevels|() &	Draws a level diagram plot\\
\verb|showPlot|() &	Shows a level diagram plot\\
\hline
\end{tabular}
\end{table*}

\begin{table*}[ht]
\caption{Method listing of  \texttt{calculations\_atom\_single.StarkMap}(atom) class}
\centering
\footnotesize
\begin{tabular}{l l}
\hline
Name (parameters) & Short description\\
\hline
\verb|defineBasis|(n, l, j, mj, nMin, ...) &	Initializes basis of states around state of interest\\
\verb|diagonalise|(eFieldList[, ...]) &	Finds atom eigenstates in a given electric field\\
\verb|plotLevelDiagram|([units, ...]) &	Makes a plot of a stark map of energy levels\\
\verb|showPlot|([interactive]) &	Shows plot made by \verb|plotLevelDiagram| \\
\verb|savePlot|([filename]) &	Saves plot made by \verb|plotLevelDiagram|\\
\verb|exportData|(fileBase[, exportFormat]) &	Exports \verb|StarkMap| calculation data\\
\verb|getPolarizability|([maxField, ...]) &	Returns the polarizability of the state (MHz~cm$^2$/V$^2$) \\
\hline
\end{tabular}
\end{table*}

\begin{table*}[ht]
\caption{Method listing of  \texttt{calculations\_atom\_pairstate.PairStateInteractions}(atom, n, l, j, nn, ll, jj, m1, m2, interactionsUpTo=1) class that calculates Rydberg level diagram (spaghetti) for the given pair-state.}
\centering
\footnotesize
\begin{tabular}{l l}
\hline
Name (parameters) & Short description\\
\hline
\verb|defineBasis|(theta, ...) &	Finds relevant states in the vicinity of the given pair-state\\
\verb|getC6perturbatively|(...) &	Calculates C6 from second order pertubation theory (GHz~$\mu$m$^6$)\\
\verb|getLeRoyRadius|() & Returns Le Roy radius for initial pair-state ($\mu$m)\\
\verb|diagonalise|(rangeR, ...) &	Finds eigenstates in atom pair basis\\
\verb|plotLevelDiagram|([...]) &	Plots pair-state level diagram\\
\verb|showPlot|([interactive]) &	Shows level diagram printed by \verb|plotLevelDiagram|\\
\verb|exportData|(fileBase[, ...]) &	Exports \verb|PairStateInteractions| calculation data\\
\verb|getC6fromLevelDiagram|(...) & 	Finds $C_6$ coefficient for original pair-state (GHz~$\mu$m$^6$).\\
\verb|getC3fromLevelDiagram|(...) & 	Finds $C_3$ coefficient for original pair-state (GHz~$\mu$m$^3$).\\
\verb|getVdwFromLevelDiagram|(...) &	Finds $r_{\rm vdW}$ coefficient for original pair-state ($\mu$m).\\
\hline
\end{tabular}
\end{table*}

\begin{table*}[ht]
\caption{Method listing of  \texttt{calculations\_atom\_pairstate.StarkMapResonances}(atom1, state1, atom2, state2) class that calculates pair-state Stark maps for finding resonances.}
\centering
\footnotesize
\begin{tabular}{l l}
\hline
Name (parameters) & Short description\\
\hline
\verb|findResonances|(nMin, ...) &	Finds near-resonant dipole-coupled pair-states\\
\verb|showPlot|([interactive]) &	Plots initial state Stark map and its dipole-coupled resonances\\
\hline
\end{tabular}
\end{table*}

\begin{table*}[ht]
\caption{Function and class listing of  \texttt{wigner} module providing support for angular element calculations}
\centering
\footnotesize
\begin{tabular}{l l}
\hline
Name (parameters) & Short description\\
\hline
\verb|Wigner3j|(j1,j2, ...) &	returns Winger 3j-coefficient\\
\verb|Wigner6j|(j1,j2,...) &	returns Wigner 6j-coefficent\\
\verb|wignerDmatrix|(theta,phi) & Class for obtaining Wigner D-matrix\\
\hline
\end{tabular}
\end{table*}

\end{document}